\documentclass[12pt]{iopart}

\usepackage{graphicx}
\usepackage[numbers]{natbib}
\usepackage{epsfig}
\usepackage{rotate}
\usepackage{epstopdf}
\usepackage{color}
\usepackage{citesort}
\usepackage{subfigure}
\usepackage{dcolumn}
\usepackage{bm}
\usepackage{array}
\usepackage{placeins}
\usepackage{amssymb}
\begin{document}

\newsavebox{\rotbox}

\def\empty{}


\def\StrichA{ --------- }
\def\StrichB{ -- -- -- -- }
\def\StrichC{ --$\,\cdot\,$--$\,\cdot\,$--$\,\cdot$ }
\def\StrichD{ ---$\,\cdot\,$---$\,\cdot$ }
\def\StrichE{ --$\,\cdot\,\cdot\,$--$\,\cdot\,\cdot$ }
\def\StrichF{ $\cdot$ $\cdot$ $\cdot$ $\cdot$ $\cdot$ $\cdot$ }
\def\StrichG{ -- $\,\cdot\,$ -- $\,\cdot$ }


\def\SymbolA{ $\Box$ }
\def\SymbolB{ {\protect\scriptsize $\bigcirc$} }
\def\SymbolC{ {\protect\small $\triangle$} }
\def\SymbolD{ + }
\def\SymbolE{ $\times$ }
\def\SymbolF{ $\Diamond$ }
\def\SymbolG{ $\Box$ }

\def\bild #1#2#3 {
\leavevmode
\def\test{#2}
\ifx\test\empty \else \epsfxsize=#2 \fi
\def\test{#3}
\ifx\test\empty \else \epsfysize=#3 \fi \hfil\epsffile{#1}\hfil }

\def\qrbild #1#2#3 {
\savebox{\rotbox}{
\leavevmode
\def\test{#2}
\ifx\test\empty \else \epsfysize=#2 \fi
\def\test{#3}
\ifx\test\empty \else \epsfxsize=#3 \fi \epsffile{#1} }
\hfil\rotr{\rotbox}\hfil }

\def\qlbild #1#2#3 {
\savebox{\rotbox}{
\leavevmode
\def\test{#2}
\ifx\test\empty \else \epsfysize=#2 \fi
\def\test{#3}
\ifx\test\empty \else \epsfxsize=#3 \fi \epsffile{#1} }
\hfil\rotl{\rotbox}\hfil }
\def\PollyStandard #1#2#3#4 {
\begin{figure}[h]
\unitlength1.0truecm
 \begin{picture}(11.0,6.9)
 \makebox(11.0,6.9)[bl]{
  \put(-0.4,-0.5){\qrbild{#1}{}{9.0truecm} }
  \savebox{\rotbox}{\large #3}
  \put( 0.3 , 6.0 ){\makebox(0,0)[t]{\rotl{\rotbox}}}
  \put( 9.0 ,-0.1 ){\makebox(0,0)[r]{\large #2}}
  \put( 9.3 , 6.3 ){\makebox(0,0)[r]{#4}}
  }
 \end{picture}\par
\end{figure}
}

\newcommand\Nu{\nobreak\mbox{$\mathcal N$\hskip-0.95mm$u$}}
\newcommand\Ra{\nobreak\mbox{$\mathcal R$\hskip-0.3mm$a$}}
\newcommand\Reyy{\nobreak\mbox{$\mathcal R$\hskip-0.3mm$e$}}
\newcommand\Pran{\nobreak\mbox{$\mathcal P$\hskip-0.3mm$r$}}
\newcommand\x{{\bf x}}

\title[Boundary layer structure in turbulent thermal convection]
{\large Boundary layer structure in turbulent thermal convection and its
consequences for the  required numerical resolution}

\author[Olga Shishkina, Richard J. A. M. Stevens, Siegfried Grossmann and Detlef Lohse]
{Olga Shishkina$^1$, Richard J. A. M. Stevens$^2$, Siegfried Grossmann$^3$, Detlef Lohse$^2$}
\vskip2mm
\address{$^1$DLR - Institute for Aerodynamics and Flow Technology,
Bunsenstra\ss e 10, D-37073 G\"ottingen, Germany}
\vskip-1.5mm
\ead{Olga.Shishkina@dlr.de}
\vskip2mm
\address{$^2$Department of Science and Technology, Impact Institute, and J.M. Burgers Center for
Fluid Dynamics, University of Twente, P.O. Box 217,  7500 AE Enschede, The Netherlands}
\vskip-1.5mm
\ead{R.J.A.M.Stevens@tnw.utwente.nl; d.lohse@utwente.nl}
\vskip2mm
\address{$^3$Fachbereich Physik der Philipps-Universit\"at, Renthof 6, D-35032 Marburg, Germany}
\vskip-1.5mm
\ead{grossmann@physik.uni-marburg.de}

\begin{abstract}
Results on the Prandtl--Blasius type kinetic and thermal boundary layer thicknesses 
in turbulent Rayleigh--B\'enard convection in a broad range of Prandtl numbers
are presented. 
By solving the laminar Prandtl--Blasius boundary layer equations, we 
calculate   
the ratio of the thermal and kinetic boundary layer thicknesses, which 
depends on the Prandtl number $\Pran$ only.
It is approximated as $0.588\Pran^{-1/2}$ for $\Pran\ll\Pran^*$
and as  $0.982\Pran^{-1/3}$ for $\Pran^*\ll\Pran$, with $\Pran^*\equiv0.046$.
Comparison of the 
Prandtl--Blasius velocity boundary layer thickness with that evaluated 
in the direct  numerical simulations
by Stevens, Verzicco, and Lohse
({\it J. Fluid Mech.} 643, 495 (2010)) gives very good agreement.
Based on the Prandtl--Blasius type considerations, 
we derive a lower-bound estimate for the minimum number of the computational mesh nodes, required to conduct accurate numerical 
simulations of moderately high (boundary layer dominated) turbulent Rayleigh--B\'enard convection, 
in the thermal and kinetic boundary layers close to bottom and top  plates.
It is shown that the number of required nodes 
within each boundary layer depends on $\Nu$ and $\Pran$ and grows with the Rayleigh number $\Ra$ not slower than $\sim\Ra^{0.15}$. 
This estimate agrees excellently with empirical results, which were 
based on the convergence of the Nusselt number in numerical simulations.
\end{abstract}

\maketitle

\section{Introduction}

Rayleigh--B\'enard (RB) convection is the classical system to study properties of thermal convection. In this system a layer of fluid confined between two horizontal plates is heated from below and cooled from above. Thermally driven flows are of utmost importance in industrial applications and in natural phenomena. Examples include the thermal convection in the atmosphere, the ocean, in buildings, in process technology, and in metal-production processes. In the geophysical and astrophysical context one may think of
  convection in Earth's mantle, in Earth's outer core, and in the outer layer of the
 Sun. 
E.g., the random reversals of Earth's or the Sun's 
magnetic field have been connected with thermal convection.

Major progress in the understanding of the Rayleigh--B\'enard system has been made over the last decades, see e.g.\ the recent reviews \cite{ahl09,loh10}. 
Meanwhile it has been well established that the general heat transfer properties of the system, i.~e. $\Nu = \Nu(\Ra,\Pran)$ and $\Reyy = \Reyy(\Nu,\Pran)$, are well described by the 
Grossmann--Lohse (GL) theory \cite{gro00,gro01,gro02,gro04}. In that
 theory, in order to estimate the thicknesses of the kinetic and thermal boundary layers (BL) and the viscous and thermal dissipation rates, the boundary layer flow is considered to be scalingwise laminar Prandtl--Blasius flow over a plate.
We use the conventional definitions: The Rayleigh number is $\Ra = \alpha g H^3 \Delta / \nu \kappa$ with the isobaric thermal expansion coefficient $\alpha$, the gravitational acceleration $g$, the height $H$ of the RB system, the temperature difference $\Delta$ between the heated lower plate and the cooled upper plate, and the material constants
 $\nu$, kinematic viscosity, and $\kappa$, thermal diffusivity, both considered to be
 constant in the container (Oberbeck--Boussinesq approximation).
The Prandtl number is defined as $\Pran=\nu/\kappa$ and the Reynolds number $\Reyy=UH/\nu$, with the wind amplitude $U$ 
which forms in  the bulk of the RB container.

The assumption of a laminar boundary layer will break down
 if the shear Reynolds number $\Reyy_s$ in the BLs becomes
 larger than approximately $420$ \cite{ll87}. Most experiments and direct numerical 
simulations (DNS) currently available are in regimes
 where the boundary layers are expected to be  still (scalingwise) 
laminar, see \cite{ahl09}. Indeed, experiments 
have confirmed that the boundary layers 
scalingwise behave as in laminar flow \cite{sun08}, i.e., follow the
{\it scaling predictions} of the Prandtl--Blasius theory 
\cite{pra04,bla08,poh21a,mek61,sch79,ll87}. 
Recently, Zhou {\it et al.}\ 
\cite{zho10,zho10b} have shown that not only the scaling of the thickness, but 
also the experimental and numerical 
{\it boundary layer profiles} in Rayleigh--B\'enard 
convection agree perfectly with the Prandtl--Blasius profiles, 
if they are evaluated in the time dependent reference frames, based on the respective {\em momentary} thicknesses. This confirms that the Prandtl--Blasius boundary layer theory is indeed the relevant theory to describe the boundary layer dynamics in Rayleigh--B\'enard convection for not too large $\Reyy_s$.

The aim of this paper is to explore the consequences of 
 the Prandtl--Blasius theory for the required numerical grid resolution of the BLs in DNSs. 
Hitherto, convergence checks can only be done {\it a posteriori}, by checking whether
the Nusselt number does not considerably change with increasing  grid resolution 
\cite{ker96,ver97,ker00,shi07,shi08,ste10} or by
guaranteeing (e.g.\ in ref.\ \cite{cal05,ste10}) that 
the Nusselt numbers calculated from the global 
energy dissipation rate or thermal dissipation rate well agree with that one calculated from 
the temperature gradient at the plates or the ones obtained from the overall heat flux. 
The knowledge that the profiles are of Prandtl--Blasius type offers the opportunity 
to {\it a priori} determine the number of required grid points in the BLs for 
given Rayleigh number and Prandtl number, valid in the boundary layer dominated ranges of moderately high $\Ra$ numbers. 

In section 2 we will first revisit the Prandtl--Blasius BL theory -- see 
 refs.\ 
\cite{pra04,bla08,poh21a,mek61,sch79,ll87} 
or for more recent discussions in the
context of RB refs.\ \cite{gro04,ahl06} -- and 
derive the ratio between the thermal boundary layer thickness ${\delta_{\theta}}$ and the velocity boundary layer thickness 
${\delta_u}$ as functions of the Prandtl number $\Pran$ extending previous work (section 3). 
We will also discuss the limiting cases for large and small $\Pran$, respectively. 
The transitional Prandtl number between the two limiting regimes turns out to be
surprisingly small, namely $\Pran^* = 0.046$. The crossover range is found to be rather broad, roughly four orders of magnitude in $\Pran$. 
In section 4 we note that the Prandtl--Blasius velocity BL  thickness 
is different from the velocity BL thickness 
based on the position of the maximum r.m.s.\ velocity fluctuations
(widely used in the literature), but well agrees 
with a BL thickness based on the position of the maximum of an energy dissipation
derivate that was recently introduced in ref.\ \cite{ste10a,ste10}. 
We then  derive the estimate for the minimum number of grid points 
 that should be placed in the boundary layers close the top and bottom 
plates, in order to guarantee proper grid resolution. 
Remarkably, the number of grid points that must have a distance smaller than $\delta_u$ from the
wall {\it increases} with increasing $\Ra$, roughly as $\sim\Ra^{0.15}$. 
This estimate is  compared with  {\it a posteriori results}  for the required grid resolution
obtained in various  DNSs of the last three decades, finding good agreement. 
Section 5 is left to conclusions.

\section{Prandtl boundary layer equations}

The Prandtl--Blasius boundary layer equations
for the velocity field ${\bf u} (x,z)$ (assumed to be two-dimensional and stationary)
 over a semi-infinite horizontal plate \cite{pra04,bla08,poh21a,mek61,sch79,ll87} read
\begin{eqnarray}
\label{prandtlequ}
 u_x\partial_x u_x+u_z\partial_z u_x=\nu\partial_z\partial_z u_x,
\end{eqnarray}
with the boundary conditions $u_x(x,0)=0$, $u_z(x,0)=0$, and $u_x(x,\infty)=U$.
Here $u_x(x,z)$ is the horizontal component of the velocity (in the direction $x$ of the large-scale circulation), $u_z(x,z)$ the vertical component of the velocity (in the direction $z$ perpendicular to the plate), and $U$ the horizontal velocity outside the kinetic boundary layer (wind of turbulence). 
Correspondingly, 
the equation determining the (stationary) temperature field $T(x,z)$ reads
\begin{eqnarray}
\label{Tequation}
 u_x\partial_x T+u_z\partial_z T=\kappa\partial_z\partial_z T,
\end{eqnarray}
with the boundary conditions $T(x,0)=T_{plate}$ and $T(x,\infty)=T_{bulk}$, which 
under Oberbeck--Boussinesq conditions is the arithmetic mean of the upper and lower plate
temperature. Applying these equations to RB flow implies that we assume the temperature
field to be passive.

The dimensionless similarity variable $\xi$ for the vertical distance $z$ from the plate measured at the distance $x$ from the plate's edge is
\begin{eqnarray}
\label{0T5}
\xi=z\sqrt{\frac{U}{x\nu}}.
\end{eqnarray}
Since the flow in Prandtl theory is two-dimensional, a streamfunction $\hat\Psi$ can be introduced, 
which represents  the velocity field.
The streamfunction is non-dimensionalized as $\Psi={\hat\Psi}/{\sqrt{x\nu U}}$, and the temperature is measured in terms of $\Delta/2$, giving
the non-dimensional temperature field $\Theta$. Rewriting eqs.\ (\ref{prandtlequ}) and
 (\ref{Tequation}) 
in terms of $\Psi$ and $\Theta$ one obtains
\begin{eqnarray}
\label{PB1}
d^3\Psi/d\xi^3+0.5\,\Psi\, d^2\Psi/d\xi^2&=&0,\\
\label{PB2}
d^2\Theta/d\xi^2+0.5\,\Pran\,\Psi\, d\Theta/d\xi&=&0.
\end{eqnarray}
Here the boundary conditions are
\begin{eqnarray}
\label{PB3}
\Psi(0)=0,&\quad d\Psi/d\xi(0)=0, \quad& d\Psi/d\xi(\infty)=1,\\
\label{PB4}
\Theta(0)=0,&& \Theta(\infty)=1.
\end{eqnarray}

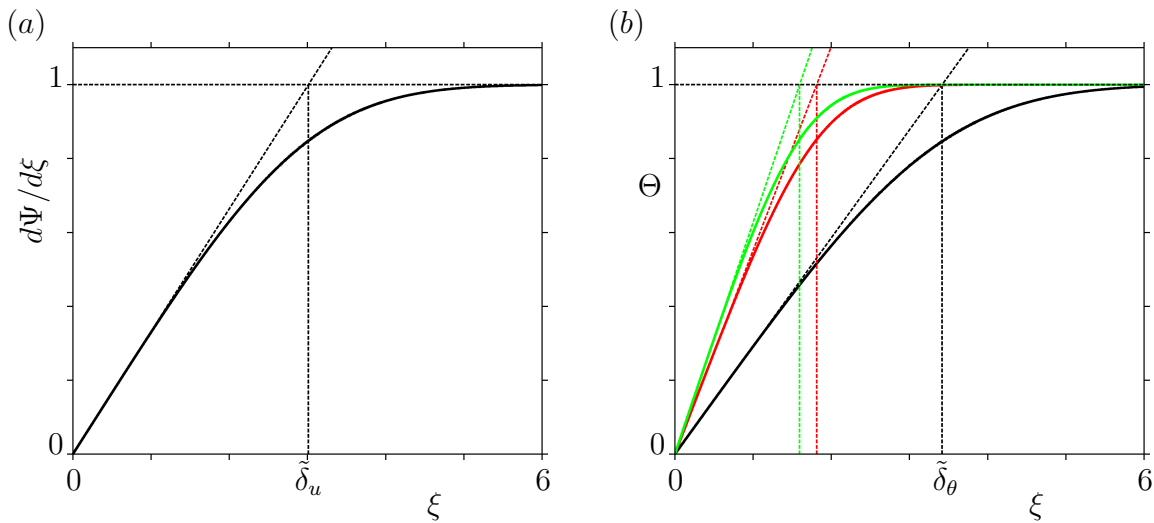
\begin{figure}
\unitlength0.8truecm
\begin{picture}(18.0,9)
\put(0.5,0.9){\bild{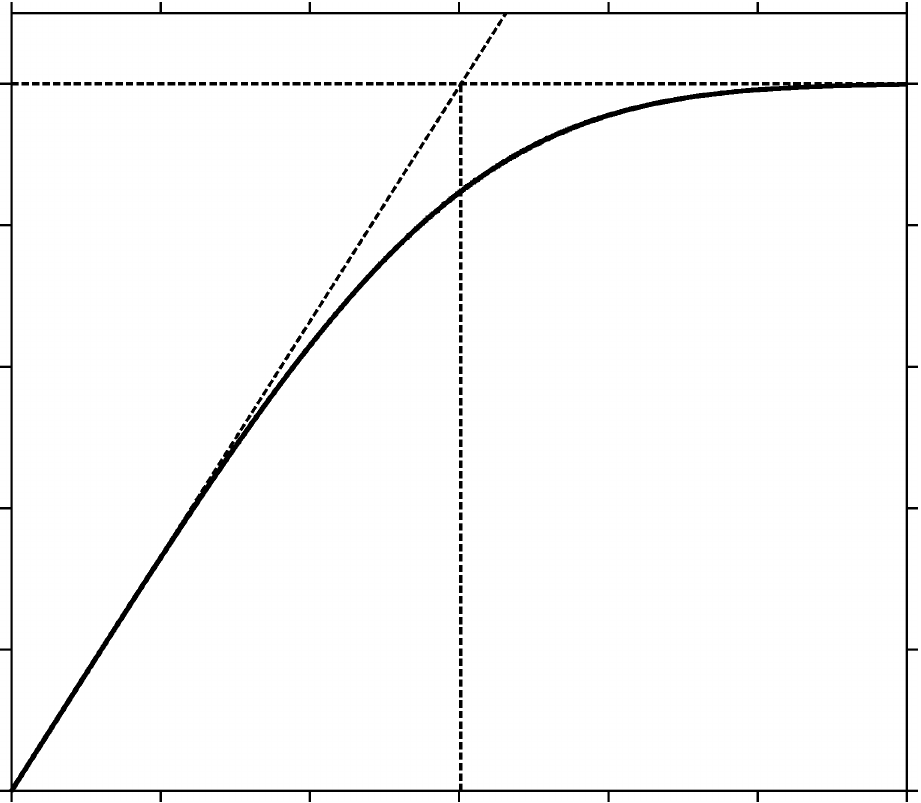}{6.4truecm}{} }
\put(10.5,0.9){\bild{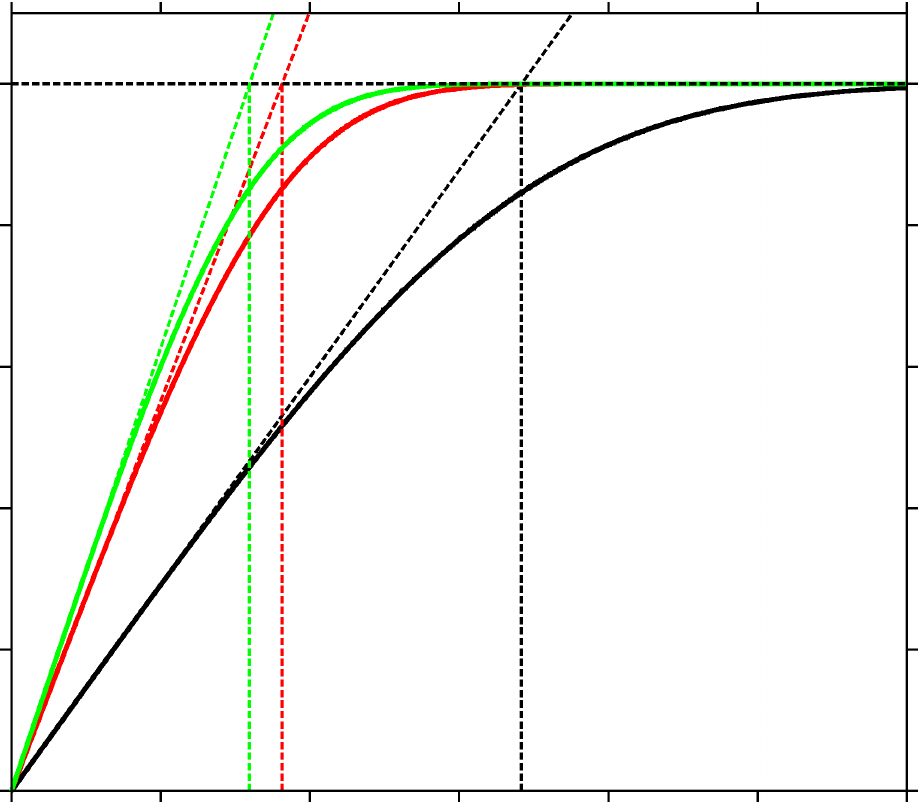}{6.4truecm}{} }
\put(1.0,0.4){0}
\put(4.8,0.4){$\tilde \delta_u$}
\put(8.8,0.4){6}
\put(7,0){$\xi$}
\put(0.7,1){0}
\put(0.7,7.1){1}
\put(0.2,4.5){\rotatebox{90}{$d\Psi/d\xi$}}
\put(0,8){$(a)$}
\put(11,0.4){0}
\put(15.4,0.4){$\tilde \delta_{\theta}$}
\put(18.8,0.4){6}
\put(17,0){$\xi$}
\put(10.7,1){0}
\put(10.7,7.1){1}
\put(10.5,5.3){$\Theta$}
\put(10,8){$(b)$}
\end{picture}
\caption{
Solution of the Prandtl--Blasius  equations (\ref{PB1})--(\ref{PB4}):
$(a)$ Longitudinal velocity profile $\frac{d\Psi}{d\xi}(\xi)$ (solid curve) with respect to the similarity variable $\xi$.
The tangent to the longitudinal velocity profile at the plate ($\xi=0$) and the straight line $d\Psi/d\xi=1$ (both dashed lines) intersect 
at $\xi=\tilde{\delta}_u\equiv A^{-1}\approx3.012$, for all $\Pran$. We define this value $\tilde{\delta}_u$ as the thickness of the kinetic boundary layer. 
$(b)$ Temperature profile $\Theta(\xi)$ as function of the similarity variable $\xi$ for $\Pran=0.7$ (black solid curve),
$\Pran=4.38$ (red solid curve) and $\Pran=6.4$ (green solid curve). The tangents to the profile curves at the plate ($\xi=0$) and the straight line $\Theta=1$ (dashed lines) define the edges (thicknesses) of the corresponding thermal boundary layers,
i.~e., $\xi=\tilde{\delta}_{\theta}\equiv C(\Pran)$. For the presented cases $\Pran=0.7$, 4.38, and 6.4 one has $C(0.7)\approx3.417$,
$C(4.38)\approx1.814$, and $C(6.4)\approx1.596$, respectively.}
\label{PIC0}
\end{figure}

The temperature and velocity profiles obtained from  numerically solving equations (\ref{PB1})--(\ref{PB4}) (for particular Prandtl numbers) are already shown in textbooks \cite{mek61,ll87,sch79} and in the context
of RB convection in refs.\ \cite{ahl06,shi09}:
From the momentum equation (\ref{PB3}) with above boundary conditions
one immediately obtains the horizontal velocity $d\Psi/d\xi$.
The dimensionless kinetic boundary layer thickness 
$\tilde{\delta}_u$ can be defined as that distance from the plate at which
 the tangent to the function $d\Psi/d\xi$ at the plate ($\xi=0$) intersects the straight line $d\Psi/d\xi=1$ (see figure~\ref{PIC0} $a$).
As equation (\ref{PB1}) and  the boundary conditions (\ref{PB3}) contain 
no parameter whatsoever,  the {\it dimensionless} thickness $\tilde{\delta}_u$ 
of the kinetic boundary layer with respect to the similarity variable $\xi$ 
is universal, i.e., independent of  $\Pran$ and $U$ or $\Reyy$,  
\begin{eqnarray}
\label{1}
\tilde{\delta}_u=A^{-1} \approx 3.012  ~~~\mbox{or} ~~~~A\approx0.332.
\end{eqnarray}

Solving numerically equation (\ref{PB2}) with the boundary conditions (\ref{PB4}) for any fixed Prandtl number,
one obtains the temperature profile with respect to the similarity variable $\xi$ (see figure~\ref{PIC0} $b$).
Note that in contrast to the longitudinal velocity $d\Psi/d\xi$, the temperature profile $\Theta$ 
depends not only on $\xi$ but also on the Prandtl number, since $\Pran$ appears in equation (\ref{PB2}) as the (only) parameter.
The distance from the plate at which the tangent to the $\Theta$ profile intersects the straight line $\Theta=1$
defines the dimensionless thickness of the thermal boundary layer, 
\begin{eqnarray}
\label{1ZV}
\tilde{\delta}_{\theta}=C(\Pran),
\end{eqnarray}
where $C(\Pran)$ is a certain function of Prandtl number. E.g., 
one  numerically finds $C\approx3.417$, 1.814, and 1.596 for
$\Pran=0.7$, 4.38, and 6.4, respectively (see figure~\ref{PIC0} $b$).

From (\ref{1}) and (\ref{1ZV}) one obtains the ratio between the (dimensional) thermal 
boundary layer thickness $\delta_{\theta}$
and the (dimensional) kinetic boundary layer thickness $\delta_{u}$:
\begin{eqnarray}
\label{2T3}
\frac{\delta_{\theta}}{\delta_u}  = \frac{\tilde{\delta}_{\theta}}{\tilde{\delta}_u} = AC(\Pran).
\end{eqnarray}
As discussed above, the constant $A$ and the function $C=C(\Pran)$ are found from the solutions
of equations (\ref{PB1})--(\ref{PB4}) for different $\Pran$. $A$ and $C(\Pran)$ reflect the slopes of the respective profiles, 
\begin{eqnarray}
\label{1T5}
A=\frac{d^2\Psi}{d\xi^2}(0), ~~~~C(\Pran)=\left[\frac{d\Theta}{d\xi}(0)\right]^{-1}.
\end{eqnarray}

With (\ref{0T5}) the physical thicknesses are $\delta_u = \tilde{\delta}_u / \sqrt{\frac{U}{x\nu}}$ and $\delta_{\theta} = \tilde{\delta}_{\theta}/\sqrt{\frac{U}{x\nu}}$, generally depending on $U$ and the position $x$ along the plate. The  {\em physical} thermal BL thickness then is 
\begin{eqnarray}
\label{thermalthickness}
\delta_{\theta} = \frac{C(\Pran)}{\sqrt{{U}/({x\nu})}} = \left[{\sqrt{\frac{U}{x\nu}} \frac{\partial \Theta}{\partial \xi} (0)}\right]^{-1} = \left[{\frac{\partial \Theta}{\partial z}(0)}\right]^{-1}.
\end{eqnarray}
Thus, explicitly it depends neither on $U$ nor on the position $x$ along the plate.
Reminding the definition of the thermal current $J = \langle u_z T \rangle - \kappa \partial_z \langle T \rangle$, we get $\langle \frac{\partial \Theta}{\partial z} (0) \rangle= \frac{1}{\Delta / 2} \langle \frac{\partial T}{\partial z} (0) \rangle = \frac{2}{\kappa \Delta} J = 2 H^{-1} \Nu$, i.~e., on $x$-average we have 
\begin{eqnarray}
\label{thermalslopethickness}
\delta_{\theta} = \frac{H}{2 \Nu}. 
\end{eqnarray}
$\delta_{\theta}$ is the so-called slope thickness, see Sect.\ 2.4  of reference \cite{ahl06}. 
In contrast to the thermal BL thickness $\delta_{\theta}$ the physical velocity BL thickness $\delta_u = A^{-1} / \sqrt{\frac{U}{x\nu}}$ depends explicitly both on the position $x$ and on the wind amplitude $U$. In a Rayleigh--B\'enard cell we choose for $x$ a representative value $x = \tilde{a} L = \tilde{a} \Gamma H$. Then the famous Prandtl formula \cite{pra04} results
\begin{eqnarray}
\label{velocitythickness}
\delta_u = \frac{aH}{\sqrt{\Reyy}}.  
\end{eqnarray}
Here $a = \sqrt{\frac{\tilde{a} \Gamma}{A^2}} = A^{-1}\sqrt{\tilde{a} \Gamma}$. The constant $a$ has been obtained empirically \cite{gro02}, based on the experimental measurements by \cite{qiu01a} performed in a cylindrical cell of aspect ratio one, filled with water. The result was \cite{gro02}
\begin{eqnarray}
\label{aaa}
a\approx0.482.
\end{eqnarray}
We note that this value probably depends on the aspect ratio, on the shape of the RB container, and can also be different for numerical 2D Rayleigh--B\'enard convection \cite{delu90,sch04,sug09}. It will also be different for the slope thickness as considered here or other definitions as e.~g.\ 
the 99\% -thickness.

It seems worthwhile to note that similarly to the case of  
$\delta_{\theta}$ also $\delta_u$ can be expressed by a profile slope at  
the plate. Analogously to the temperature case one calculates for the  
kinetic thickness $\delta_u = U / \frac{\partial u_x}{\partial z}(0)$.  
Here $U$ appears explicitly and the derivative may depend on $\x$. The  
denominator is the local stress tensor component, which -- after  
averaging -- describes the momentum transport, just as the temperature  
profile derivative at the plate characterises the heat transport. In  
combination with eq.~(\ref{velocitythickness}) it says that the kinetic stress behaves as  
$\langle \frac{\partial u_x}{\partial z} (0) \rangle \sim  U \sqrt{\Reyy} / (a H)$.

\begin{figure}
\unitlength0.8truecm
\begin{picture}(18.0,9)
\put(4.45,1.1){\bild{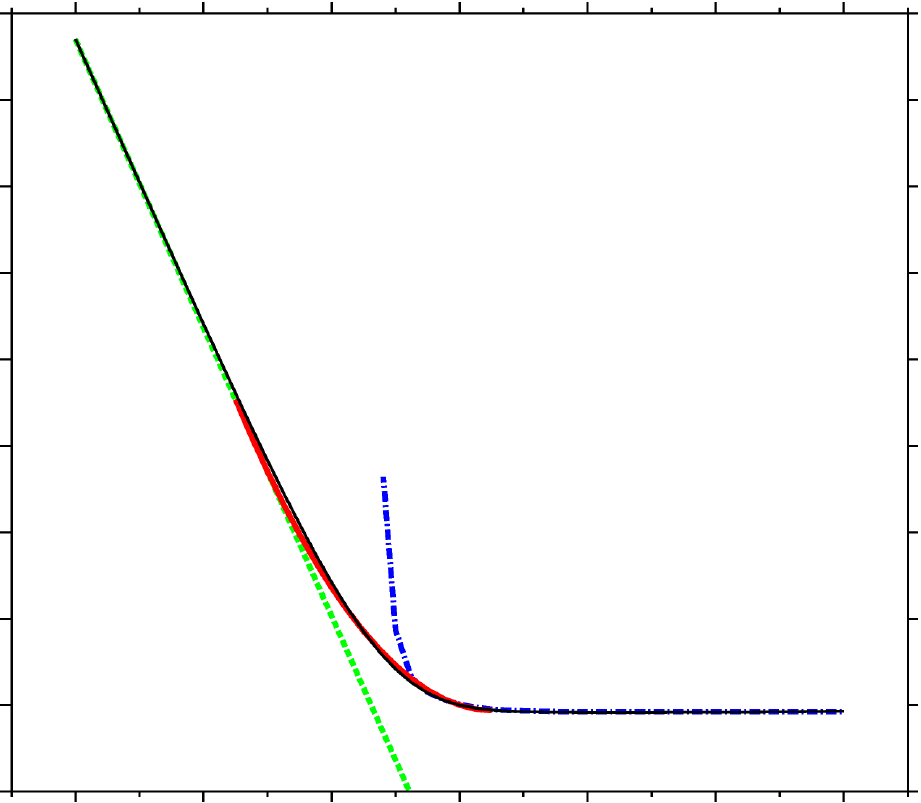}{6.45truecm}{} }
\put(4.8,0.6){-7}
\put(5.365,0.6){-6}
\put(5.93,0.6){-5}
\put(6.495,0.6){-4}
\put(7.06,0.6){-3}
\put(7.625,0.6){-2}
\put(8.19,0.6){-1}
\put(8.855,0.6){0}
\put(9.42,0.6){1}
\put(9.985,0.6){2}
\put(10.55,0.6){3}
\put(11.115,0.6){4}
\put(11.68,0.6){5}
\put(12.245,0.6){6}
\put(12.81,0.6){7}
\put(10.5,0){$\log\Pran$}
\put(4.1,1.15){-0.1}
\put(4.25,1.9){0.0}
\put(4.25,2.65){0.1}
\put(4.25,3.4){0.2}
\put(4.25,4.15){0.3}
\put(4.25,4.9){0.4}
\put(4.25,5.65){0.5}
\put(4.25,6.4){0.6}
\put(4.25,7.15){0.7}
\put(4.25,7.9){0.8}
\put(2.9,4.0){\rotatebox{90}{$\log\left[({\delta_{\theta}}/{\delta_u})\Pran^{1/3}\right]$}}
\end{picture}
\caption{Double-logarithmic plot of the ratio of the thermal and kinetic boundary layer thicknesses, normalised by $\Pran^{-1/3}$, as obtained from numerical solution of equations (\ref{PB1})--(\ref{PB4}) as function of  $\Pran$ (solid 
black line). For large $\Pran$ the curve through the data is constant, for small $\Pran$ the (plotted, reduced) curve behaves $\propto \Pran^{-1/6}$. Approximation (\ref{5}) (green dotted line) is indistinguishable from  ${\delta_{\theta}}/{\delta_u}$ in the region $\Pran<3\times10^{-4}$. Approximation (\ref{13}) (blue dashed-dotted line) well represents  ${\delta_{\theta}}/{\delta_u}$ for $\Pran>0.3$; for $\Pran>3$ it practically coincides with approximation (\ref{13T5}). Approximation (\ref{14})
(red solid curve) connects the analytical approximations in the transition
 range $3\times10^{-4}\leq\Pran\leq3$ between the lower and upper Prandtl number regimes.}
\label{PIC1}
\end{figure}
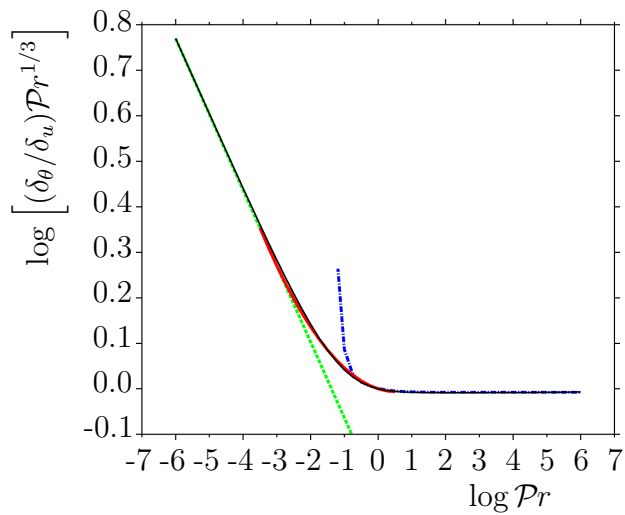

From eqs.\ (\ref{2T3}) and (\ref{velocitythickness}) we also find the useful (and known) relation
for Prandtl-Blasius boundary layers
\begin{eqnarray}
\label{thermalkinetic}
\delta_{\theta} = a_{\theta} C(\Pran) \frac{H}{\sqrt{\Reyy}}  ~~~~\mbox{with}~~~ a_{\theta} = A\cdot a \approx 0.160.
\end{eqnarray}
From solving equations 
(\ref{PB1})--(\ref{PB4}) together with  relations  (\ref{1T5}) one obtains that 
the BL thickness ratio   (\ref{2T3}) 
has two limiting cases, namely  
${\delta_{\theta}}/{\delta_u}\sim\Pran^{-1/2}$ for very small $\Pran\ll1$ and
${\delta_{\theta}}/{\delta_u}\sim\Pran^{-1/3}$ for very large $\Pran\gg1$. We thus 
 present the ratio of the thermal and kinetic boundary layer thicknesses normalised by $\Pran^{-1/3}$ in figure~\ref{PIC1} for different $\Pran$ from $\Pran=10^{-6}$ to $10^{6}$.
The figure confirms that the scaling of the ratio between the thermal and kinetic boundary layer thicknesses in the low and high Prandtl number regimes is $\Pran^{-1/2}$ and $\Pran^{-1/3}$, respectively. Between these two limiting regimes there is a transition region, whose width is about 4 orders of magnitude in $\Pran$. In the next section we will derive analytic expressions for the ratio ${\delta_{\theta}}/{\delta_u}$ in the respective regimes, which will be used in the remainder of the paper to 
analyse the resolution  properties for DNS in the BLs of the Rayleigh--B\'enard system.

In the Prandtl--Blasius theory the asymptotic velocity amplitude $U$ is a given parameter; 
the resulting heat current $\Nu$ is a performance of the boundary layers only. In contrast, in Rayleigh--B\'enard convection the heat transport is determined by the BLs together with the bulk flow. Therefore in RB convection the wind amplitude $U$ no longer is a passive parameter, but $U$ and $\Nu$ are actively coupled properties of the full thermal convection process.   

The Reynolds number $\Reyy$ is  defined as the dimensionless wind amplitude, 
\begin{eqnarray}
\label{2-1}
 \Reyy=\frac{UH}{\nu}.
\end{eqnarray}
From the law for the kinetic BL thickness (\ref{velocitythickness}), the thermal BL thickness $\delta_\theta$ (\ref{thermalslopethickness}), and the BL thickness ratio 
(\ref{2T3}) one obtains
\begin{eqnarray}
\label{rere}
 \Reyy&=&\left(\frac{aH}{\delta_u}\right)^2=\left(\frac{\delta_\theta}{\delta_u}\right)^2
\left(\frac{aH}{\delta_\theta}\right)^2=
4a^2\Nu^2\left(\frac{\delta_\theta}{\delta_u}\right)^2.
\end{eqnarray}
This $\Reyy \sim  \Nu^2$ law is in perfect agreement with the GL theory \cite{gro00,gro01,gro02,gro04}. {In that theory several sub-regimes in the $(\Ra,\Pran)$ parameter space are introduced, depending on the dominance of the BL or bulk contributions. In regimes I and II the BL of the temperature field dominates, while in III and VI it is the thermal bulk. Regimes I and II differ in the velocity field contributions: It either is the $u$-BL (I) or the $u$-bulk (II) which dominates; analogously the pair III and IV is characterized. The labels $\ell$ 
(for {\bf l}ower Pr) and $u$ (for {\bf u}pper Pr) distinguish the cases in which the thermal BL is thicker or smaller than the kinetic one. All ranges in the GL theory, which are thermal boundary layer dominated, show the  $\Reyy \sim  \Nu^2$ behaviour, namely $I_l$, $I_u$, $II_l$, $II_u$. In the thermal bulk dominated ranges of RB convection the relation between $\Reyy$ and $\Nu$ is different. } 
In $III_u$ we have $\Reyy \sim \Nu^{4/3}$, in $IV_l$ it is $\Reyy \sim \Nu$, and in $IV_u$
also $\Reyy \sim  \Nu^{4/3}$ holds; but here the Prandtl--Blasius result (\ref{rere}) is not applicable, 
since the heat transport mainly depends on the heat transport properties of the bulk. 
In the range $I_{\infty}$, although boundary layer dominated, also 
a different relation ($\Reyy \sim \Nu^3$) 
holds; here the upper and the lower kinetic BLs fill the whole volume and 
therefore there is no free flow outside the BLs, in contrast to the Prandtl--Blasius assumption
of an asymptotic velocity with the LSC amplitude $U$.

\section{Approximations for the ratio ${\delta_{\theta}}/{\delta_u}$ of the temperature and velocity boundary layer thicknesses} 

In this section we will derive analytical approximations for the ratio ${\delta_{\theta}}/{\delta_u}$ for the three regimes identified in the previous section, cf. figure~\ref{PIC1}. We start by discussing the low ($\Pran < 3\times10^{-4}$) and the high  ($3<\Pran$)  Prandtl number regimes, before we discuss the transition region $3\times10^{-4}\leq\Pran\leq3$. 

\subsection{Approximation of ${\delta_{\theta}}/{\delta_u}$ for $\Pran<3\times10^{-4}$}

In the case of very small Prandtl number, $\Pran\ll1$, the thickness of the velocity boundary layer is negligible compared with the thickness of the temperature boundary layer, i.e., ${\delta_{\theta}}\gg{\delta_u}$. Hence, in most of the thermal boundary layer it is ${u_x}\approx U$. Introducing the similarity variable as in ref.~\cite{sch79}
\begin{eqnarray}
\label{2T7}
\eta=\frac{z}{2}\sqrt{\frac{U}{x\kappa}},
\end{eqnarray}
one obtains the following equation for the temperature as a function of $\eta$:
\begin{eqnarray*}
d^2\Theta/d\eta^2+2\eta \,d\Theta/d\eta=0, ~~~\mbox{with} ~~~~ 
\Theta(0)=0, \quad\Theta(\infty)=1.
\end{eqnarray*}
The solution of this boundary value problem is the Gaussian error function
\begin{eqnarray}
\label{3}
\Theta(\eta)=\mbox{erf} (\eta)\equiv\frac{2}{\sqrt{\pi}}\int_0^\eta e^{-t^2}dt.
\end{eqnarray}
According to (\ref{0T5}) and (\ref{2T7}), the similarity variable $\xi$ used in the Prandtl equations and the similarity variable $\eta$ used in the approximation for $\Pran\ll1$ are related as follows
\begin{eqnarray}
\label{4}
\eta=\frac{1}{2}\Pran^{1/2}\xi.
\end{eqnarray}
Applying now the formulae (\ref{3}), (\ref{4}) and (\ref{1T5}) we obtain the following 
 equalities:
\begin{eqnarray*}
\frac{2}{\sqrt{\pi}}=\frac{d\Theta}{d\eta}(0)=\frac{d\Theta}{d\xi}(0)\cdot\frac{d\xi}{d\eta}=\frac{1}{C(\Pran)}\cdot2\Pran^{-1/2}.
\end{eqnarray*}
This leads to the approximation for the function $C(\Pran)=\sqrt{\pi}\Pran^{-1/2}$ for very small $\Pran$.
\begin{eqnarray}
\label{5}
\frac{\delta_{\theta}}{\delta_u}=A\sqrt{\pi}\Pran^{-1/2}\approx0.588\Pran^{-1/2}, ~~~~ \Pran\ll1.
\end{eqnarray}

In figure~\ref{PIC1} one can see that for very small Prandtl numbers, $\Pran<3\times10^{-4}$, the approximation (\ref{5}) is as expected indistinguishable 
from the numerically obtained ${\delta_{\theta}}/{\delta_u}$.

\subsection{Approximation of ${\delta_{\theta}}/{\delta_u}$ for $\Pran>3$}

Meksyn \cite{mek61}, based on the work by Pohlhausen \cite{poh21a}, derived that the solution of the temperature equation (\ref{PB2}), together with relation (\ref{PB4}) equals 
\begin{eqnarray}
\label{tempsolution}
\Theta\left(\frac{\xi}{\sqrt{2}}\right)=D\int_0^{{\xi}/{\sqrt{2}}}e^{-F(t)\Pran}dt,\qquad
F(t)=\frac{1}{\sqrt{2}}\int_0^t {\Psi}(q)dq.
\end{eqnarray}
The constant $D$ can be found as usual from the boundary condition at infinity
and was approximated in \cite{poh21a,mek61} for $\Pran>1$ as follows
\begin{eqnarray*}
D=\frac{0.478\Pran^{1/3}}{c(\Pran)},\qquad
c(\Pran)\approx1+\frac{1}{45\Pran}-\frac{1}{405\Pran^2}+\frac{161}{601425\Pran^3}-...
\end{eqnarray*}
From this and (\ref{tempsolution}) one derives 
\begin{eqnarray*}
\frac{0.478\Pran^{1/3}}{c(\Pran)} = D=
\frac{d\Theta}{d({\xi}/{\sqrt{2}})}(0)=
\sqrt{2}\frac{d\Theta}{d\xi}(0)=
\frac{\sqrt{2}}{C(\Pran)}.
\end{eqnarray*}
This connects $c(\Pran)$ and $C(\Pran)$ as follows
\begin{eqnarray*}
C(\Pran)\approx\frac{\sqrt{2}}{0.478}\,c(\Pran)\Pran^{-1/3}\approx2.959\,c(\Pran)\Pran^{-1/3},
\end{eqnarray*}
resulting in the approximation 
\begin{eqnarray}
\label{13}
\frac{\delta_{\theta}}{\delta_u}= AC(\Pran)= E\Pran^{-1/3}c(\Pran),\quad E\approx A\frac{\sqrt{2}}{0.478}\approx0.982.
\end{eqnarray}
For $\Pr \gg 1$, 
the function $c(\Pran)$ approaches 1, hence $C(\Pran)\approx 2.959\Pran^{-1/3}$, implying
\begin{eqnarray}
\label{13T5}
\frac{\delta_{\theta}}{\delta_u}&=&E\Pran^{-1/3}, ~~~~ \Pran\gg1.
\end{eqnarray}
In figure~\ref{PIC1} the approximation (\ref{13}) is presented as a blue dash-dotted curve.
For $\Pran>3$ the function $({\delta_{\theta}}/{\delta_u)\Pran^{1/3}}$ almost coincides with the constant $E$.

\subsection{Approximation of ${\delta_{\theta}}/{\delta_u}$ in the crossover range $3\times10^{-4}\leq\Pran\leq3$}

As one can see in figure~\ref{PIC1}, the approximation (\ref{5}) well represents ${\delta_{\theta}}/{\delta_u}$ in the region $\Pran<3\times10^{-4}$, while 
 (\ref{13T5}) is a good approximation of ${\delta_{\theta}}/{\delta_u}$ for $\Pran>3$. An approximation of the ratio of the thermal and kinetic boundary layer thicknesses in the transition region $3\times10^{-4}\leq\Pran\leq3$ is obtained by applying a least square fit to the numerical solutions of the Prandtl--Blasius equations (\ref{PB1})-(\ref{PB4}). One finds:
\begin{eqnarray}
\label{14}
\frac{\delta_{\theta}}{\delta_u}&\approx&\Pran^{-0.357+0.022\log\Pran}, ~~~~~ 3\times10^{-4}\leq\Pran\leq3.
\end{eqnarray}
As seen in figure~\ref{PIC1}, this relation is a good fit of the full 
solution in the transition regime.

\subsection{Summary}

For the ratio $\delta_{\theta}/ \delta_u$ of the thicknesses of the thermal and kinetic boundary layers, which depends strongly (and only) on $\Pran$, we find according to (\ref{5}), (\ref{13T5}), and (\ref{14})
\begin{eqnarray}
\label{Res1}
\frac{\delta_{\theta}}{\delta_u}=
\left\{\begin{tabular}{lcl}
           $A\sqrt{\pi}\Pran^{-1/2},\;A\approx0.332$,&&$\Pran<3\times10^{-4}$,\\
           $\Pran^{-0.357+0.022\log\Pran},$&&$3\times10^{-4}\leq\Pran\leq3,$\\
           $E\,\Pran^{-1/3},\;E\approx0.982$,&&$\Pran>3$.
      \end{tabular} 
\right.
\end{eqnarray}
The crossover Prandtl number $\Pran^*$ between the asymptotic behaviours, cf. first and last line of (\ref{Res1}), is defined as the intersection point $\Pran^* = 0.046$ of the asymptotic approximations. Note that this  crossover between the small-$\Pran$ behaviour $\delta_{\theta}/\delta_u \propto \Pran^{-1/2}$ and the large-$\Pran$ behaviour $\delta_{\theta}/\delta_u \propto \Pran^{-1/3}$ does not
happen at a Prandtl number of order 1, but at the  
more than 20 times smaller value $\Pran^* = 0.046$. 
In this sense most experiments are conducted in the large $\Pran$ regime.  However, also note
that other definitions of the BL thicknesses lead to other crossover Prandtl numbers.

Finally, we also give the thickness of the kinetic BL in the three regimes, 
as obtained from (\ref{Res1}) and (\ref{thermalslopethickness}), namely
\begin{eqnarray}
\label{visc}
{\delta_u}=
\left\{\begin{tabular}{ll}
           $ 0.5\Nu^{-1}\Pran^{1/2}A^{-1}\pi^{-1/2}H$,        &$\Pran<3\times10^{-4}$,\\
           $ 0.5\Nu^{-1}\Pran^{0.357-0.022 \log \Pran}H$, &$3\times10^{-4}\leq \Pran \leq3$,\\
           $ 0.5\Nu^{-1}\Pran^{1/3}E^{-1}H$,                &$\Pran>3$. 
      \end{tabular} \right.
\end{eqnarray}

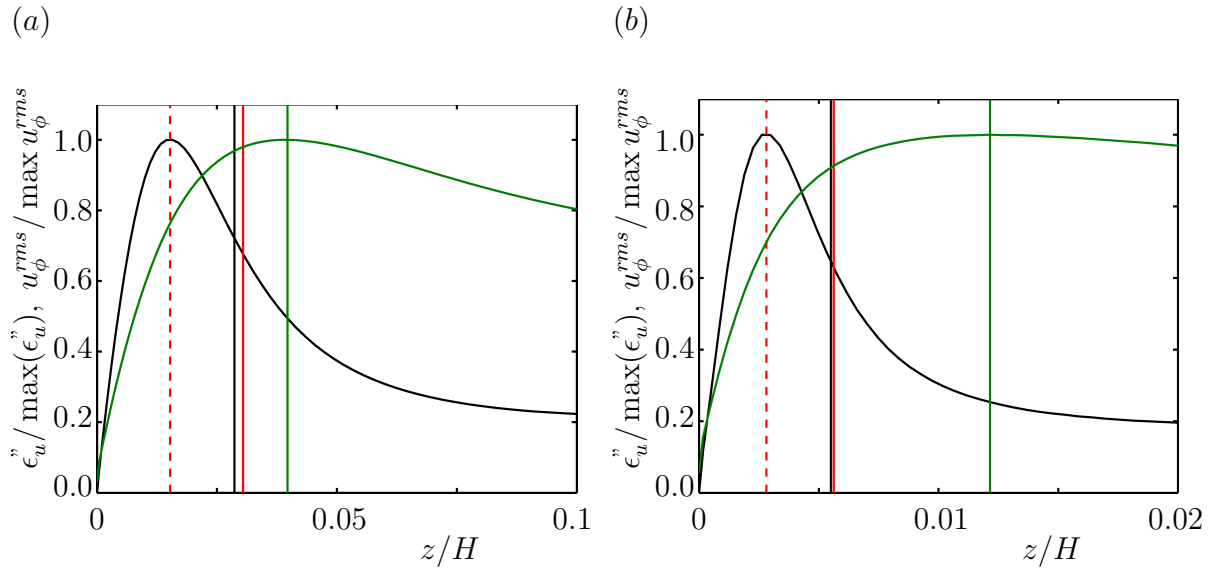
\begin{figure}
\unitlength0.8truecm
\begin{picture}(18.0,9.5)
\put(0.9,1.1){\bild{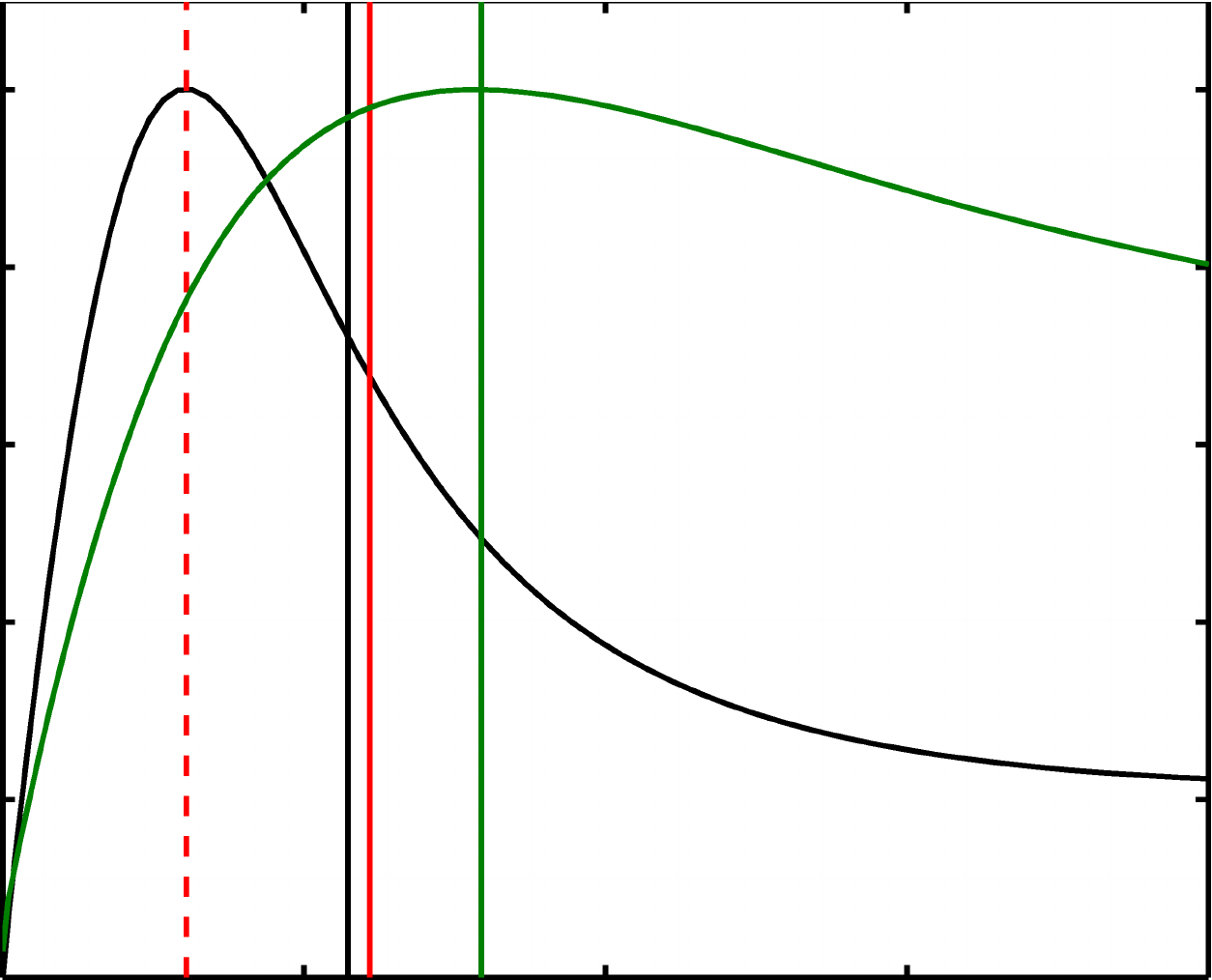}{6.4truecm}{} }
\put(1.3,0.5){0}
\put(5.0,0.5){0.05}
\put(9.0,0.5){0.1}
\put(6.8,0){$z/H$}
\put(0.7,1.05){0.0}
\put(0.7,2.2){0.2}
\put(0.7,3.35){0.4}
\put(0.7,4.5){0.6}
\put(0.7,5.65){0.8}
\put(0.7,6.8){1.0}
\put(0,1.5){\rotatebox{90}{$\epsilon_u^"/\max(\epsilon_u^"),\; u_\phi^{rms}/\max{u_\phi^{rms}}$}}
\put(0,8.8){$(a)$}
\put(10.9,1.1){\bild{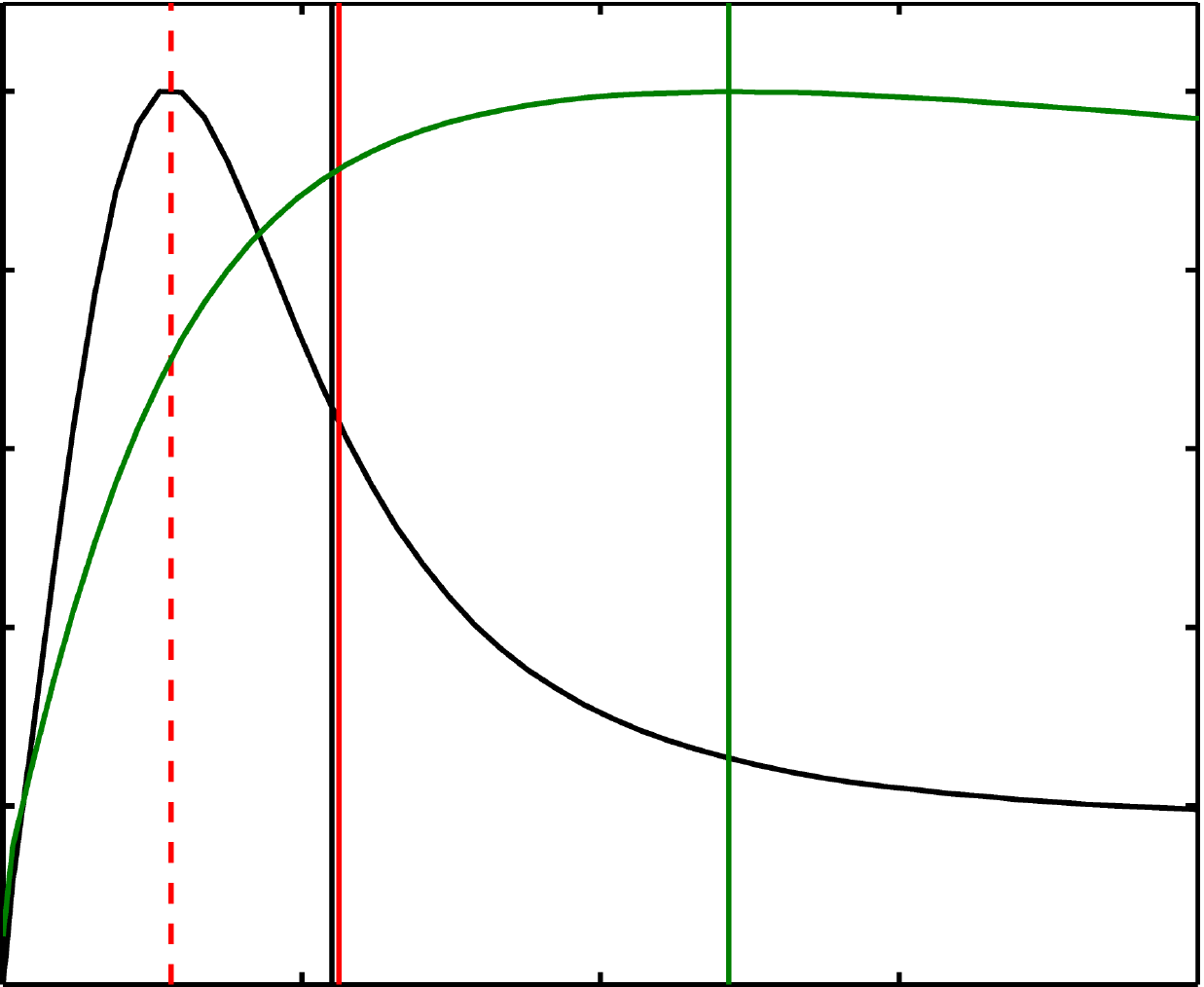}{6.4truecm}{} }
\put(11.3,0.5){0}
\put(15.0,0.5){0.01}
\put(18.9,0.5){0.02}
\put(16.8,0){$z/H$}
\put(10.7,1.05){0.0}
\put(10.7,2.2){0.2}
\put(10.7,3.35){0.4}
\put(10.7,4.5){0.6}
\put(10.7,5.65){0.8}
\put(10.7,6.8){1.0}
\put(10,1.5){\rotatebox{90}{$\epsilon_u^"/\max(\epsilon_u^"),\; u_\phi^{rms}/\max{u_\phi^{rms}}$}}
\put(10,8.8){$(b)$}
\end{picture}
\caption{Profiles of $\epsilon_u^"$ (\ref{eps}) (black), and the r.m.s. velocity fluctuations for the azimuthal velocity component $u_\phi$ (green) for $(a)$ $\Ra=10^8$ and $\Pran=6.4$ and $(b)$ $\Ra=2\times10^9$ and $\Pran=0.7$. The profiles have been normalised with the respective maxima for clarity. The vertical black lines indicate the velocity boundary layer thickness based on (\ref{visc}). The red dashed and solid lines indicate the heights at which the quantity $\epsilon_u^"$ (\ref{eps}) 
has a maximum  and two times this height, respectively. The vertical green line indicates the position of the maximum r.m.s. velocity fluctuations.}
\label{Figure kineticBLthickness}
\end{figure}

We compare this Prandtl--Blasius  result (\ref{visc}) for the kinetic boundary layer thickness in terms of $\Nu$ and $\Pran$ 
(thus valid if the heat transport is BL dominated) 
with the estimate given in reference\ \cite{ste10}, where the kinetic boundary layer thickness 
in a cylindrical cell is identified as two times that height at which the averaged quantity
\begin{eqnarray}
\label{eps}
\epsilon_u^":= \langle {\bf u} \cdot \nabla^2 {\bf u} \rangle_{t,\phi,r}
\end{eqnarray}
has a maximum, because it was empirically found that the maximum of $\epsilon_u^"$  is approximately in the middle of the velocity boundary layer. Here $\bf{u}$ is the velocity field and the averaging is over time $t$, the azimuthal direction $\phi$, and over the radial direction $0.1R <r< 0.9R$, with $R$ the radius of the cylindrical convective cell. The restricted
range for the radial direction has been used in order to exclude the singularity region close to the cylinder axis and the region close to the sidewall, where the definition misrepresents the kinetic boundary layer thickness.  Figure \ref{Figure kineticBLthickness} shows that there is a very good agreement between the theoretical Prandtl--Blasius slope
boundary layer thickness and that obtained using (\ref{eps}). 
The figure also shows that the position of the maximum r.m.s. velocity fluctuations is not a good indicator for the velocity boundary layer edge; it rather seems to identify the position where the LSC is the strongest.

\section{Resolution requirements within the boundary layers in DNS}

We now come to the main point of the paper: What can we learn from the Prandtl--Blasius theory
for the required mesh resolution in the BLs of DNS of turbulent RB convection? 
Obviously, a 
``proper'' mesh resolution should be used 
in order to obtain accurate results.  In a perfect DNS the local 
mesh size should be smaller than the  local 
Kolmogorov $\eta_K (\x,t) $ and Batchelor $\eta_B (\x,t) $ 
scales (see e.g.\ ref.\ \cite{my75}), and the 
resolution in the boundary layers should be also sufficient, see e.g.\ 
 \cite{gro83,ker96,shi09,shi09b,ste10}.
It indeed has been well established that the Nusselt number is very sensitive to the grid resolution used in the boundary layers;  when DNS is underresolved, the measured Nusselt number is
 too high \cite{gro83,ver03,ker96,ama05,ver08,ver08b,ste10}. 
Hitherto, the standard way to empirically check whether 
the mesh resolution is sufficient is to try a finer mesh and to make sure that the
Nusselt number is not too different. In this way the minimal number of grid points 
that is needed in the boundary layer is obtained by trial and error: 
Gr\"otzbach \cite{gro83} varied the number of grid points in the boundary layer between $1$ and $5$ in simulations up to $\Ra=3\times10^5$ with $\Pran=0.71$ and found that $3$ grid points in the boundary layers should be sufficient. Verzicco and Camussi 
\cite{ver03} tested this at $\Ra=2\times10^7$ and $\Pran=0.7$ and 
stated that at least $5$ points should be placed in the boundary layers. 
Stevens {\it et al.}\ \cite{ste10} 
tested the grid resolution for $\Ra=2\times10^6$ to $2\times10^{11}$ 
and $\Pran=0.7$. 
They found  that for $\Ra=2\times10^9$ the minimum number 
of nodes in the boundary layers 
should be around  $10$ 
and that this number increases for increasing $\Ra$.
Together with the earlier  series of papers the data clearly
 suggest that indeed there is an increase of required grid points in the BL with
increasing Rayleigh number. 

However, one must be careful. The empirical determination of the required number of grid points in the BL is not only intensive in 
computational cost, but also difficult. The Nusselt number obtained in the simulations not only depends on the grid resolution 
in the BLs at the top and bottom plates, but  also on the grid resolution in the {\it bulk} and at the side walls where the thermal 
plumes pass along \cite{ste10}. So obviously a general theory-based criterion for the required grid resolution  
in the thermal and kinematic boundary layers
will be helpful for performing future simulations. 
In this section we will derive such 
a universal criterion, harvesting above results from the Prandtl--Blasius boundary layer theory.

{We first define the (local) kinetic energy dissipation rates per mass, 
\begin{eqnarray}
\label{3-0}
\epsilon_u (\x, t )  
\equiv\frac{\nu}{2}\sum_i\sum_j\left(\frac{\partial u_i (\x,t)}{\partial x_j}+\frac{\partial u_j (\x,t)}{\partial x_i}\right)^2. 
\end{eqnarray}
Its time and space average for incompressible flow with zero velocity b.c. is $\langle \epsilon_u \rangle_{t,V} = \nu\sum_i\sum_j \langle \left(\frac{\partial u_i (\x,t)}{\partial x_j}\right)^2 \rangle_{t,V} $. It is connected with the Nusselt number through the exact relation }

%
%
\begin{eqnarray}
\label{3-1}
\langle\epsilon_u\rangle_{t,V}=\frac{\nu^3}{H^4}(\Nu-1)\Ra\Pran^{-2}.
\end{eqnarray}
This follows directly from the momentum  equation
 for Rayleigh--B\'enard convection in Boussinesq approximation \cite{sig94}.
Here, 
$\langle\cdot\rangle_{t,V}$ 
denotes averaging 
over the whole volume of the convective cell and over time and (later)   
$\langle\cdot\rangle_{t,A}$ denotes averaging 
 over any  horizontal plane  and time.

We start with the well established criterion that in a perfect DNS simulation the (local) mesh
 size must not be larger than  the  (local) 
Kolmogorov scale \cite{kol41} $\eta_K (\x ,t )$,
which is  locally defined with the energy dissipation rate of the velocity,
\begin{equation}
\label{S}
\eta_K (\x , t) = \left({\nu^3}/{\epsilon_u (\x,t)}\right)^{1/4}. 
\end{equation}
$\eta_K$ is the length scale at which the inertial term $\sim u_r^2/r$ and the viscous term
$\sim \nu u_r/r^2$ of the Navier-Stokes equation balance, where $u_r \sim (\epsilon_u r)^{1/3}$
has been assumed for the velocity difference at scale $r$. A corresponding length scale $\eta_T$ follows
from the balance of the advection term $\sim u_r T_r/r$ and the thermal diffusion term $\kappa  T_r/r^2$
in the advection equation; it is
\begin{equation}
\label{S_T}
\eta_T (\x ,t) =\left({\kappa^3}/{\epsilon_u (\x , t)}\right)^{1/4} = \eta_K  (\x ,t ) \Pran^{-3/4}.
\end{equation}
However, for large Pr the velocity field is smooth at those scales at which the temperature
field is still fluctuating. Then the velocity difference $u_r \sim \sqrt{\epsilon_u/\nu} r$ and
advection term and thermal diffusion term balance at the so-called Batchelor 
scale \cite{bat59} $\eta_B$, 
which is defined as
\begin{equation}
\label{S_B}
\eta_B (\x ,t) =\left({\nu\kappa^2}/{\epsilon_u (\x , t)}\right)^{1/4} = 
\eta_K (\x , t)  \Pran^{-1/2}.
\end{equation}
For small $\Pran < 1$ obviously 
$\eta_T >  \eta_B > \eta_K$ and for comparison with the grid resolution, the Kolmogorov scale $\eta_K$
seems to be the most restrictive (i.e., smallest) length scale.   In contrast, 
for large $\Pran > 1$ it 
$\eta_T < \eta_B < \eta_K$ and one may argue that $\eta_T$ is the most restrictive length scale.
This indeed may be the case in the Prandtl number regime in which the velocity field can still
be described through Kolmogorov scaling $u_r \sim (\epsilon_u r)^{1/3}$, but for even larger $\Pran$
the velocity field becomes smooth $u_r \sim \sqrt{\epsilon_u/\nu} r$
and then the grid resolution should be compared to 
the  Batchelor scale $\eta_B$ as smallest relevant length scale. In below analysis, for 
$\Pran > 1$  we will restrict 
ourselves to this limiting case.

We now define {\it global} Kolmogorov and Batchelor length scales
$\eta_K^{global}\equiv \frac{\nu^{3/4}}{\langle\epsilon_u\rangle^{1/4}_{t,V}} $ and
$\eta_B^{global} \equiv \frac{\nu^{1/4}\kappa^{1/2}}{\langle\epsilon_u\rangle^{1/4}_{t,V}}$,
respectively, (and also the global length scale 
$\eta_T^{global} \equiv \frac{\kappa^{3/4}}{\langle\epsilon_u\rangle^{1/4}_{t,V}}$).
Using the exact relation  (\ref{3-1}), one can find how  
the global  Kolmogorov length $\eta_K^{global}$ depends
on $\Ra$, $\Pran$, and $\Nu$, namely 
\begin{eqnarray}
\label{4-3}
\eta_K^{global} \equiv
\frac{\nu^{3/4}}{\langle\epsilon_u\rangle^{1/4}_{t,V}}=
\frac{\Pran^{1/2}}{\Ra^{1/4}(\Nu-1)^{1/4}}H.
\end{eqnarray}
The admissible global mesh size $h^{global}$ should clearly be smaller 
than both $\eta_K^{global}$ and $\eta_B^{global}$, which implies that one is on the
safe side provided that 
\begin{eqnarray}
\label{4-3-1}
h^{global}\leq
\frac{\Pran^{1/2}}{\Ra^{1/4}(\Nu-1)^{1/4}}H  
\quad \hbox{for} \quad \Pran \leq 1 
\end{eqnarray}
or with the 
relation (\ref{S_B}) between the Kolmogorov and Batchelor length 
\begin{eqnarray}
\label{4-3-11}
h^{global} \leq \frac{1}{\Ra^{1/4}(\Nu-1)^{1/4}}H 
\quad \hbox{for} \quad \Pran > 1.
\end{eqnarray}
A similar way to estimate mesh requirements in the bulk was suggested for the first time by 
Gr\"otzbach \cite{gro83}. Note that with these estimates for the required bulk resolution for most times and
locations one is 
on the safe side,
as equation (\ref{3-1}) is an estimate for the volume averaged energy dissipation rate, which is localized in the boundary layers. However, not only the background field but also plumes detaching from the boundary layers do require
an adequate resolution.

To estimate the number of nodes that should be placed in the  boundary layers, 
we will first estimate the area averaged energy dissipation rate
in a horizontal plane in the velocity BL, $\langle\epsilon_u\rangle_{t,A\in BL}$.
Employing eqs.\  (\ref{2-1}), (\ref{velocitythickness}) and (\ref{3-0}),
one can find a lower bound for this quantity, namely
\begin{eqnarray}
\label{3-2}
\langle\epsilon_u\rangle_{t,A\in BL}&\geq&  \nu\left<   \left(\frac{\partial u_x}{\partial z}\right)^2\right>_{t,A}\geq
\nu\left(\left<\frac{\partial u_x}{\partial z}\right>_{t,A}\right)^2 \approx 
\nu\left(\frac{U}{\delta_u}\right)^2=\nonumber\\
&=&\nu\left(\frac{\nu\Reyy}{H}\frac{\Reyy^{1/2}}{aH}\right)^2=\frac{\nu^3\Reyy^3}{a^2H^4}.
\end{eqnarray}
From eqs.\ (\ref{3-1}), (\ref{3-2}), (\ref{rere}) and (\ref{Res1}) it follows a lower bound
for the ratio 
\begin{eqnarray}
\label{3-3}
\frac{\langle\epsilon_u\rangle_{t,A\in BL}}{\langle\epsilon_u\rangle_{t,V}} \geq
\frac{\Pran^2\Reyy^3}{a^2\Ra\Nu}=64a^4\Nu^5\frac{\Pran^2}{\Ra}\left(\frac{\delta_\theta}{\delta_u}\right)^6\nonumber\\
\qquad=\left\{\begin{tabular}{ll}
           $64{\pi}^3a^4A^6\Nu^5\Pran^{-1}\Ra^{-1}$,&$\Pran< 3\times10^{-4}$,\\
           $64a^4\Nu^5\Pran^{-0.15+0.132\log \Pran}\Ra^{-1}$,&$3\times 10^{-4} \le \Pran \le 3$,\\
           $64a^4E^6\Nu^5\Ra^{-1}$,&$\Pran>3$.
      \end{tabular} \right.
\end{eqnarray}
For the Kolmogorov length $\eta_K^{BL}$ in the velocity BL one can therefore write
\begin{eqnarray}
\label{hs}
\eta_K^{BL} 
\equiv 
\left<\left(\frac{\nu^3}{\epsilon_u}\right)^{1/4}\right>_{t,A\in BL}
\approx 
\left(\frac{\langle\epsilon_u\rangle_{t,V}}{\langle\epsilon_u\rangle_{t,A\in BL}}\right)^{1/4}\,
\eta_K^{global}.
\end{eqnarray}
The mesh size $h^{BL}$ in the BL must be  smaller than $\eta_K^{BL}$ and $\eta_B^{BL}$,
i.e., one is on the safe side if
\begin{eqnarray}
\label{4-4}
h^{BL} \lesssim
\left\{\begin{tabular}{ll}
           $2^{-3/2}a^{-1}\Nu^{-3/2}\Pran^{3/4}A^{-3/2}\pi^{-3/4}H$, &$\Pran<3\times10^{-4}$,\\
           $2^{-3/2}a^{-1}\Nu^{-3/2}Pr^{0.5355-0.033 \log \Pran}H$,  &$3\times10^{-4}\leq \Pran \leq1$,\\
           $2^{-3/2}a^{-1}\Nu^{-3/2}Pr^{0.0355-0.033\log \Pran}H$,  &$1<\Pran\leq3$,\\
           $2^{-3/2}a^{-1}E^{-3/2}\Nu^{-3/2}H$,          &$\Pran>3$,
      \end{tabular} \right.
\end{eqnarray}
according to (\ref{3-3}), (\ref{hs}), (\ref{4-3-1}) and (\ref{4-3-11}).

From the relations (\ref{4-4}), (\ref{Res1}) and (\ref{thermalslopethickness}) one can estimate the minimum number of nodes of the computational mesh, which must be placed in each thermal and kinetic boundary layer close the plates. We find that this minimum number of nodes in the thermal boundary layers is 
\begin{eqnarray}
\label{4-5}
N_{\texttt{th.BL}}&\equiv& \frac{\delta_\theta}{h^{BL}}\nonumber\\ &\gtrsim&
\left\{\begin{tabular}{ll}
           $\sqrt{2} a \Nu^{1/2} \Pran^{-3/4} A^{3/2} \pi^{3/4}$,   &$\Pran<3\times10^{-4}$,\\
           $\sqrt{2} a \Nu^{1/2} \Pran^{-0.5355+0.033 \log \Pran}$, &$3\times10^{-4}\leq \Pran \leq1$,\\
           $\sqrt{2} a \Nu^{1/2} \Pran^{-0.0355+0.033 \log \Pran}$, &$1<\Pran\leq3$,\\
           $\sqrt{2} a \Nu^{1/2} E^{3/2}$,              &$\Pran>3$,
      \end{tabular} \right.
\end{eqnarray}
while the minimum number of nodes in the kinetic boundary layers is
\begin{eqnarray}
\label{4-6}
N_{\texttt{v.BL}}&\equiv& \frac{\delta_u}{h^{BL}}= \frac{\delta_u}{\delta_\theta} \frac{\delta_\theta}{h^{BL}}\nonumber\\ &\gtrsim&
\left\{\begin{tabular}{ll}
           $\sqrt{2}a\Nu^{1/2} \Pran^{-1/4} A^{1/2} \pi^{1/4}$,   &$\Pran<3\times10^{-4}$,\\
           $\sqrt{2}a\Nu^{1/2} \Pran^{-0.1785+0.011 \log \Pran}$, &$3\times10^{-4}\leq \Pran \leq1$,\\
           $\sqrt{2}a\Nu^{1/2} \Pran^{ 0.3215+0.011 \log \Pran}$, &$1<\Pran\leq3$,\\
           $\sqrt{2}a\Nu^{1/2} \Pran^{1/3} E^{1/2}$,             &$\Pran>3$.
      \end{tabular} \right.
\end{eqnarray}
The number of nodes in the thermal boundary layer looks very restrictive for very low $\Pran$; however, one should realise that for very low $\Pran$ the thermal boundary layer also becomes much thicker than the velocity boundary layer. Hence, the criterion for the number of nodes in the thermal boundary layers determines the ideal distribution of nodes above the viscous boundary layer. For very high $\Pran$ the kinetic boundary layer becomes much thicker than the thermal boundary layer, and hence the restriction for the velocity boundary layer determines the ideal distribution of nodes above the thermal boundary boundary layer.
Note that for large $\Pran$ equation (\ref{4-5}) suggests that the number of grid points in the
thermal boundary layer becomes independent of $\Pran$ (for fixed $\Nu$): Indeed, as the velocity
field is smooth anyhow, with increasing $\Pran$ no extra grid points are  necessary in the thermal BL.

\begin{figure} 
\unitlength1truecm 
\begin{picture}(18.0,7) 
\put(2.65,0.8){\bild{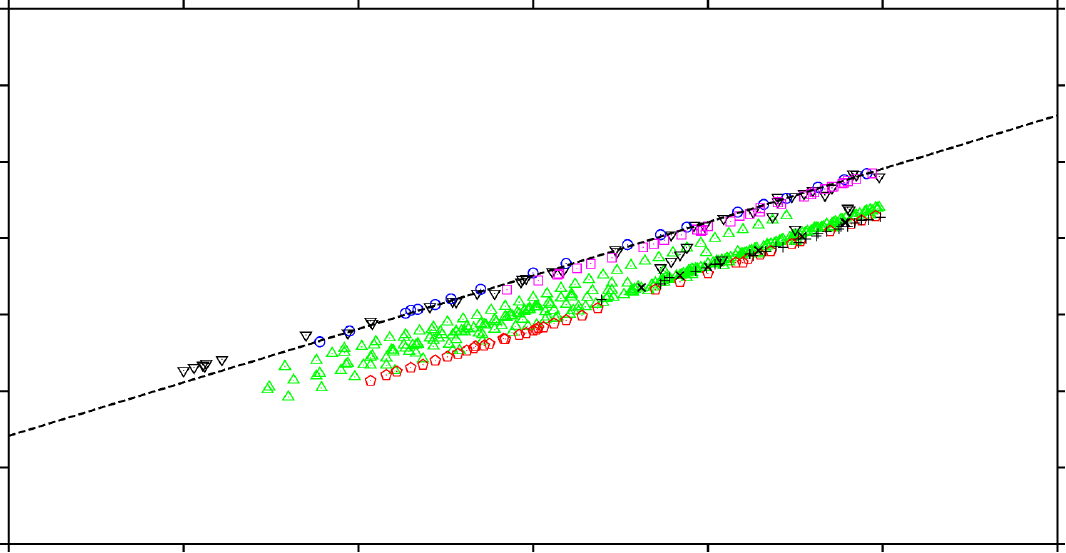}{10truecm}{} } 
\put(2.8,0.8){0} 
\put(2.5,1.5){0.2} 
\put(2.5,2.2){0.4} 
\put(2.5,2.9){0.6} 
\put(2.5,3.6){0.8} 
\put(2.5,4.4){1.0} 
\put(2.5,5.1){1.2} 
\put(2.5,5.8){1.4} 
\put(1.9,3.5){\rotatebox{90}{$\log N_{\texttt{th.BL}}$}} 
\put(1,5.6){$(a)$} 
\end{picture} 
\begin{picture}(18.0,6) 
\put(2.65,0.8){\bild{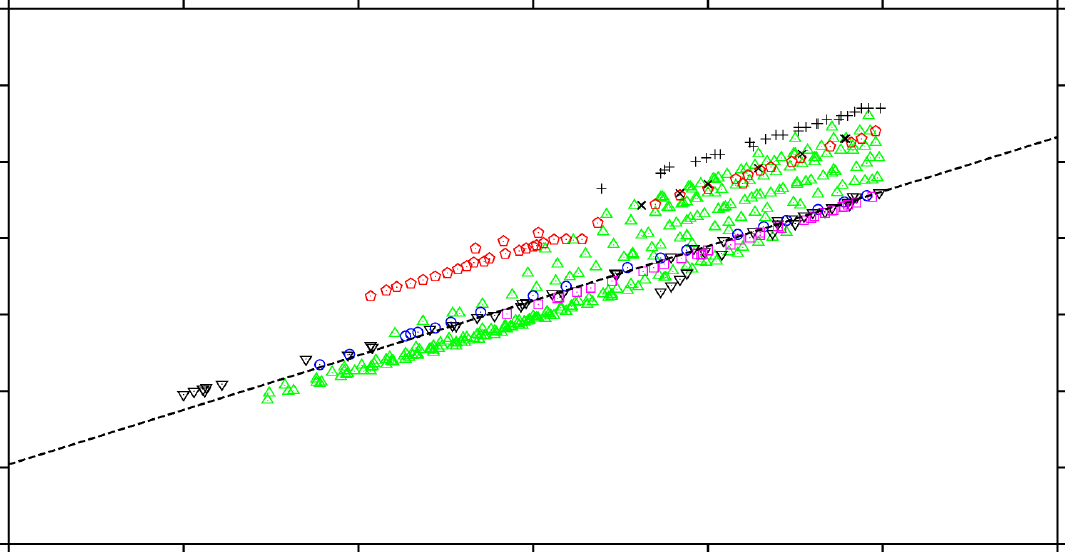}{10truecm}{} } 
\put(3.05,0.4){5} 
\put(4.65,0.4){6} 
\put(6.3,0.4){7} 
\put(7.95,0.4){8} 
\put(9.6,0.4){9} 
\put(11.1,0.4){10} 
\put(12.75,0.4){11} 
\put(8.3,0){$\log\Ra$} 
\put(2.8,0.8){0} 
\put(2.5,1.5){0.2} 
\put(2.5,2.2){0.4} 
\put(2.5,2.9){0.6} 
\put(2.5,3.6){0.8} 
\put(2.5,4.4){1.0} 
\put(2.5,5.1){1.2} 
\put(2.5,5.8){1.4} 
\put(1.9,3.5){\rotatebox{90}{$\log N_{\texttt{v.BL}}$}} 
\put(1,5.6){$(b)$} 
\end{picture} 
\caption{Minimum number of BL nodes necessary in DNS of boundary layer dominated, moderately high RB convection. $(a)$ $N_{\texttt{th.BL}}$ (\ref{4-5}) in the thermal boundary layers and $(b)$ $N_{\texttt{v.BL}}$ (\ref{4-6}) in the kinetic boundary layers, required to simulate the experimentally investigated cases, references \cite{ahl09b} (lilac squares, $\Pran=0.67$), \cite{cha01} (black triangles, $0.60\leq\Pran\leq7.00$), \cite{nie03} (blue circles, $0.68\leq\Pran\leq5.92$), \cite{roc04} (green triangles, $0.73\leq\Pran\leq6.00$), 
\cite{sun05e} (red pentagons, $3.76\leq\Pran\leq5.54$), \cite{xia02} (black crosses, $\Pran=4.2$) and \cite{qiu98} (black pluses, $\Pran=7.0$). 
Dashed lines are fits to the quasi-data (measured values introduced into eqs. (\ref{4-5}), (\ref{4-6})), with precision O($10^{-4}$); rounding the respective numbers to their upper bounds gives $(a)$ $N_{\texttt{th.BL}}\approx0.35 \Ra^{0.15}$ (\ref{4-7}) and $(b)$ $N_{\texttt{v.BL}}\approx 0.31  \Ra^{0.15}$ (\ref{4-8}) for the quasi-data in the ranges $10^6\leq\Ra\leq10^{10}$ and $0.67\leq\Pran\leq0.73$.} 
\label{PIC3} 
\end{figure}
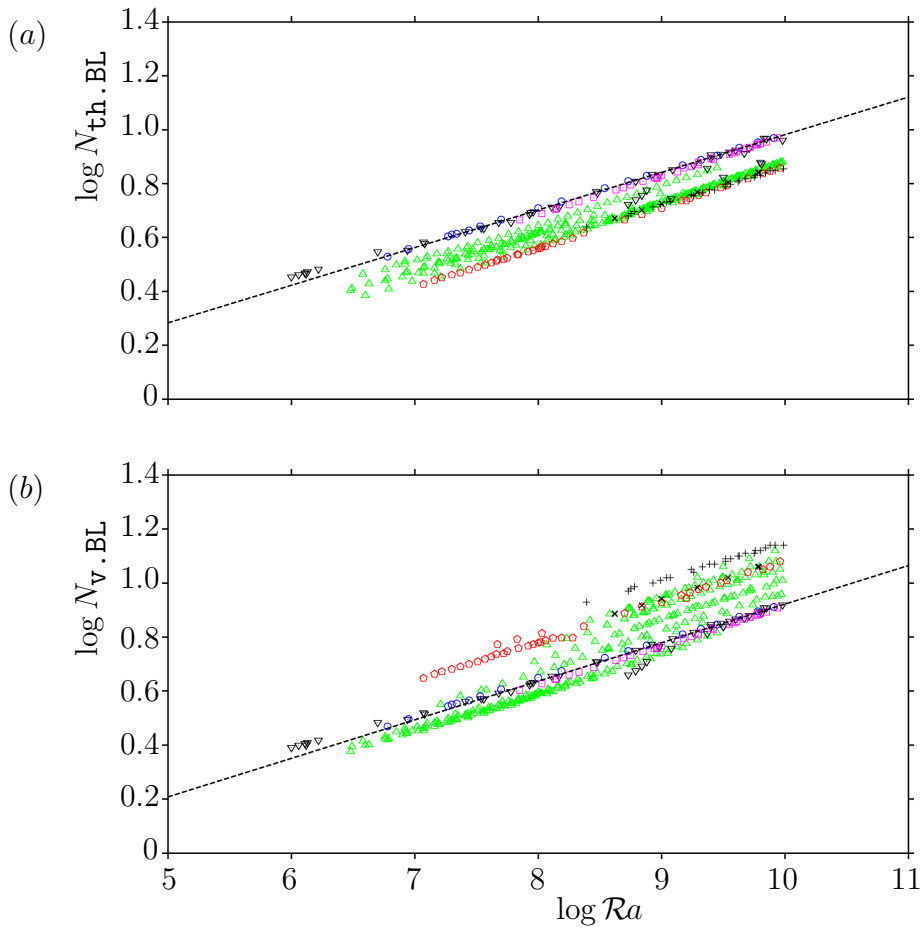

In figure~\ref{PIC3} we show the minimum number of nodes $N_{\texttt{th.BL}}$ and $N_{\texttt{v.BL}}$, respectively, necessary to simulate the cases which have been investigated experimentally so far, for different $\Ra$ and $\Pran$. The data points are generated by introducing the experimental values of $\Ra, \Pran$, and the (measured) corresponding $\Nu$ into the formulas (\ref{4-5}), (\ref{4-6}). Based on these quasi-data points,
one can give e. g. the following fits for the minimum number of nodes within the boundary layers for the case of $\Pran \approx 0.7$:
\begin{eqnarray}
\label{4-7}
N_{\texttt{th.BL}}\approx0.35\Ra^{0.15}, ~~~ 10^6 \le \Ra \le 10^{10}, \\
\label{4-8}
~N_{\texttt{v.BL}} ~ \approx0.31\Ra^{0.15}, ~~~ 10^6 \le \Ra \le 10^{10}.
\end{eqnarray}
Note that the numerical pre-factors in these estimates significantly depend on the Prandtl number and on the empirically determined (ref.~\cite{gro02}) value of $a$, cf. eq. (\ref{aaa}). The minimum node numbers for other values of $\Pran$ can be calculated directly using the relations (\ref{4-5})--(\ref{4-6}). Apparently the scaling exponent depends much less on $\Pran$. -- All these estimates only give lower bounds on the required number of nodes in the boundary layers.

As discussed at the beginning of this section, previous studies by 
Gr\"otzbach \cite{gro83}, Verzicco and Camussi \cite{ver03}, and Stevens {\it et al.}\ \cite{ste10}
found an increasing number of nodes that should be placed in the thermal and kinetic  boundary layers. The theoretical results thus confirm all above studies, because the increasing number of nodes was due to the increasing $\Ra$ number at which the tests were performed. To be more specific: according to the estimates (\ref{4-7}) and (\ref{4-8}) for $\Pran=0.7$ the minimum number of nodes that should be placed in the thermal and kinetic boundary layers is $N \approx 2.3$ for $\Ra=3\times10^5$, $N \approx 4.4$ for $\Ra=2\times 10^7$, and $N \approx 8.7$ for $\Ra=2\times10^9$. The empirically found values at the respective $\Ra$ with $\Pran \approx 0.7$ are $3$ for $\Ra=3\times10^5$, $5$ for $\Ra=2\times 10^7$, and $10$ for $\Ra=2\times10^9$. Thus there is very good agreement between the theoretical results and the empirically obtained values, especially if one considers the difficulties involved in determining these values empirically, and the empirical value for the constant $a$ (\ref{aaa}) that is used in the theoretical estimates. We want to emphasize that not only the boundary layers close to the plates, but also the kinetic boundary layers close to the vertical walls must be well resolved.

To sum up, the mesh resolution should be analysed {\it a priori} using the resolution requirements in the bulk (\ref{4-3-1}), (\ref{4-3-11}) {\it and} in the boundary layers (\ref{4-5}), (\ref{4-6}). Having conducted the DNS, the Kolmogorov and Batchelor scale should be checked {\it a posteriori}, to make sure that the mesh size was indeed small enough (as it has been done, for example, in refs.\ \cite{shi07,shi08}).

\section{Conclusion}
In summary, we used laminar Prandtl--Blasius boundary layer theory to determine the relative thicknesses of the thermal and kinetic boundary layers as functions of $\Pran$ (\ref{Res1}).

We found that neither the position of the maximum r.m.s.\ velocity fluctuations nor the position of the horizontal velocity maximum
reflect the slope velocity boundary layer thickness, although many studies use these as criteria to determine the boundary layer thickness. In contrast to them, the algorithm by Stevens 
{\it et al.}\ \cite{ste10} agrees very well with the theoretical estimate of the
kinetic slope boundary layer thickness.

We used the results obtained from the Prandtl--Blasius boundary layer theory to derive a lower bound on the minimum number of nodes that should be placed in the thermal and kinetic boundary layers close to the plates. We found that this minimum number of nodes increases not slower than $\sim\Ra^{0.15}$ with increasing $\Ra$. This result is in excellent agreement with results from several numerical studies over the last decades, in which this minimum number of nodes was determined empirically. Hence, the derived estimates can be used as guideline for future direct numerical simulations.

\ack
OS expresses her thanks to Prof. Dr.-Ing. Claus Wagner and to the Deutsche Forschungsgemeinschaft (DFG) for supporting the work under the grant WA~1510/9.  The Twente part of the work is supported by the Foundation for Fundamental Research on Matter (FOM), sponsored by NWO.


\providecommand{\newblock}{}

\end{document}